\documentclass[lettersize,journal]{IEEEtran}
\usepackage{amsmath,amsfonts}
\usepackage{algorithmic}
\usepackage{algorithm}
\usepackage{array}
\usepackage{textcomp}
\usepackage{stfloats}
\usepackage{url}
\usepackage{verbatim}
\usepackage{graphicx}
\usepackage{cite}
\usepackage{subfigure}
\usepackage{xcolor}
\newcommand{\Fig}[1]{Fig.\,{\ref{#1}}}

\begin{document}

\title{Analytical Framework to Model Reconfigurable \\ Metasurfaces including Lumped Elements}

\author{Mario Pérez-Escribano, Salvador Moreno-Rodríguez, Carlos Molero,~\IEEEmembership{Member,~IEEE,} \\ Juan F. Valenzuela-Valdés,  Pablo Padilla,  Antonio Alex-Amor
        % <-this % stops a space
\thanks{This work has been supported by grant PID2020-112545RB-C54 funded by MCIN/AEI/10.13039/501100011033 and by the European Union NextGeneration EU/PRTR. It has also been supported by grants PDC2022-133900-I00, TED2021-129938B-I00, and TED2021-131699B-I00, and by Ministerio de Universidades and the European Union NextGenerationEU, under Programa Margarita Salas, and by MCIN/AEI/10.13039/501100011033 and the European Union NextGenerationEU/PRTR under grant IJC2020-043599-I.}
\thanks{\textit{(Corresponding author: Antonio Alex-Amor)}}
\thanks{Mario Pérez-Escribano is with the Telecommunication Research Institute (TELMA), Universidad de M\'{a}laga, E.T.S. Ingenier\'{i}a de Telecomunicaci\'{o}n, 29010 Málaga, Spain.}
\thanks{Salvador Moreno-Rodríguez, Carlos Molero, Juan F. Valenzuela-Valdés, and Pablo Padilla  are with the Department of Signal Theory, Telematics and Communications, Research Centre for Information and Communication Technologies (CITIC-UGR), Universidad de Granada, 18071 Granada, Spain.}
\thanks{Antonio Alex-Amor is with the Department of Information Technologies, Universidad San Pablo-CEU, CEU Universities,  Campus Montepríncipe, 28668 Boadilla del Monte (Madrid), Spain.}

}

\markboth{}%
{M. Pérez-Escribano \MakeLowercase{\textit{et al.}}: Analytical Framework to Model Reconfigurable Metasurfaces including Lumped Elements}


\maketitle

\begin{abstract}
This paper presents an analytical framework, based on Floquet modal expansions of the electromagnetic fields and equivalent circuits, to model reconfigurable metasurfaces loaded with generic lumped elements (resistors, capacitors, inductors, varactors, etc.). The analytical approach is computationally efficient compared to full-wave solvers. Additionally, it works under oblique-incidence conditions in a wideband range of frequencies, even far beyond the onset of the first grating lobe (diffraction regime). The analytical framework is validated with some numerical examples in the commercial software CST Studio Suite, demonstrating its potential for analyzing and designing RF and microwave devices, including lumped elements, such as absorbers, polarizers, and reflectarray/transmitarray cells. 

\end{abstract}


%------------------------------------------------------------------
\section{Introduction}
\IEEEPARstart{L}{umped} elements are, by definition, circuit elements (resistors, inductors, capacitors, diodes, transistors, etc.) of much smaller size compared to the operating wavelength \cite{booklumped}. The design and application of lumped elements have been an intense research object in the radiofrequency (RF) and microwave communities in the last decades \cite{Introlumped}. Lumped elements have been traditionally combined with distributed elements, e.g., transmission lines, in integrated and planar technologies for the realization of passive and active microwave circuits such as filters, absorbers, 3-dB quadrature hybrids, impedance transformers, amplifiers, rectifiers or time-modulated antennas \cite{booklumped, Introlumped, lumpedintegrated1, lumpedintegrated2, lumpedintegrated3, JSGomez2019}. Recently, with the appearance of metamaterials, artificial composites that can go beyond the conventional properties of materials in nature, novel and sophisticated lumped-element-based microwave, and mm-wave devices have seen the light. Examples of lumped-element-based metastructures, such as periodically-loaded phase shifters or diode-tunable absorbers, reflectarray/transmitarray cells, and intelligent metasurfaces, can be found in \cite{metalumped0,  metalumped2, Tretyakovlumped}.

 %\cite{booklumped, lumpedintegrated1, lumpedintegrated2, lumpedintegrated3, lumpedintegrated4, JSGomez2019, timeresistor}.
%\cite{EnghetaBook2006}
%\cite{metalumped0, metalumped1, metalumped2, metalumped3, Tretyakovlumped}.

The study of microwave devices and metamaterials, including the presence of lumped elements, is a nontrivial task. The interaction and electromagnetic coupling between the lumped and distributed elements considerably hinder the analysis. Most of the state-of-art numerical approaches are based on full-wave schemes, such as the finite-difference time-domain (FDTD) \cite{FDTDlumped1,  FDTDlumped3} and the finite element method (FEM) \cite{FEMlumped1}. Iterative techniques \cite{iterativelumped1} and the transmission line method (TLM) \cite{TLMlumped} have also proven to be practical solutions. More recently, homogenization techniques have been applied to analyze metastructures using time-modulated lumped elements \cite{Wu2020}. 

%\cite{FDTDlumped1, FDTDlumped2, FDTDlumped3}
%\cite{FEMlumped1, FEMlumped2}
%\cite{iterativelumped1, iterativelumped2}

In general, precise analytical or quasi-analytical methods are preferred over numerical ones due to their simplicity and computational efficiency in resolving the problem \cite{Tretyakov1}. This paper proposes an alternative fully-analytical approach based on Floquet-Bloch series expansions and equivalent circuits to analyze 2D metamaterials loaded with generic lumped elements. The analytical approach is computationally efficient and gives physical insight into the scattering phenomena via the related equivalent circuit. Furthermore, it works under oblique-incidence conditions in a wideband range, even for frequency regions above the diffraction regime. 

The paper is organized as follows. Section II describes the analytical framework, as well as the advantages and limitations of the method. Section III illustrates some numerical examples to validate the approach. Section IV shows the utilization of the analytical approach for the design of a dual wideband absorber. Finally, conclusions are drawn in Section V.

%-----------------------------------------------------------------
\section{Circuit Topology and Theoretical Framework}

%%%%%%%%%%%%%%%%%%%%%%%%%%%%%%%%%%%%%%%%%%%%%%%%%%%%%%
\subsection{Circuit Topology}
\begin{figure}[!t]             
        \centering	         {\includegraphics[width= 0.8\columnwidth]{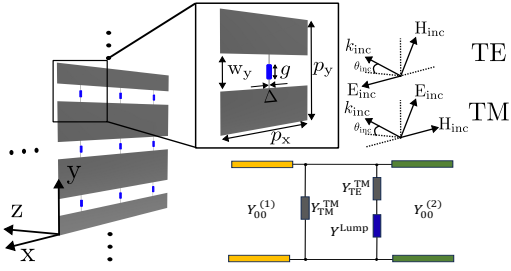}}
        \hspace*{-0.5cm}
        \caption{Sketch of the 2D metasurface loaded with lumped elements.}
        \label{Fig1}
\end{figure}

The structure under consideration consists of a metasurface fed by the incidence of an oblique (angle $\theta$) plane wave. The transverse electric-field vector is directed along the perpendicular direction of the strips to excite a richer phenomenology. It corresponds to a TE ($\phi = 0^\mathrm{o}$) or a TM ($\phi = 90^\mathrm{o}$) polarization, as shown in \Fig{Fig1}. Rows of lumped elements, such as networks based on resistors, inductors, and capacitors, will be placed periodically at the slit aperture connecting two consecutive strips. This makes the 1D grating a two-dimensional structure, as shown in \Fig{Fig1}. The distance between adjacent strips determines periodicity along the $y$ direction, whereas periodicity along the $x$ direction is defined by the distance between two consecutive lumped elements. In \Fig{Fig1}, $p_x$ and $p_y$ are the periods of the 2D grating along the $x$ and $y$ directions, 
$w_x = p_x - \Delta$ and $w_y$ are the slit widths, $\Delta$ is the thickness of the strip, and $g$ is the gap in which the lumped element is placed. 

\begin{figure}[!t]             
        \centering	           
        {\includegraphics[width=0.99\columnwidth]{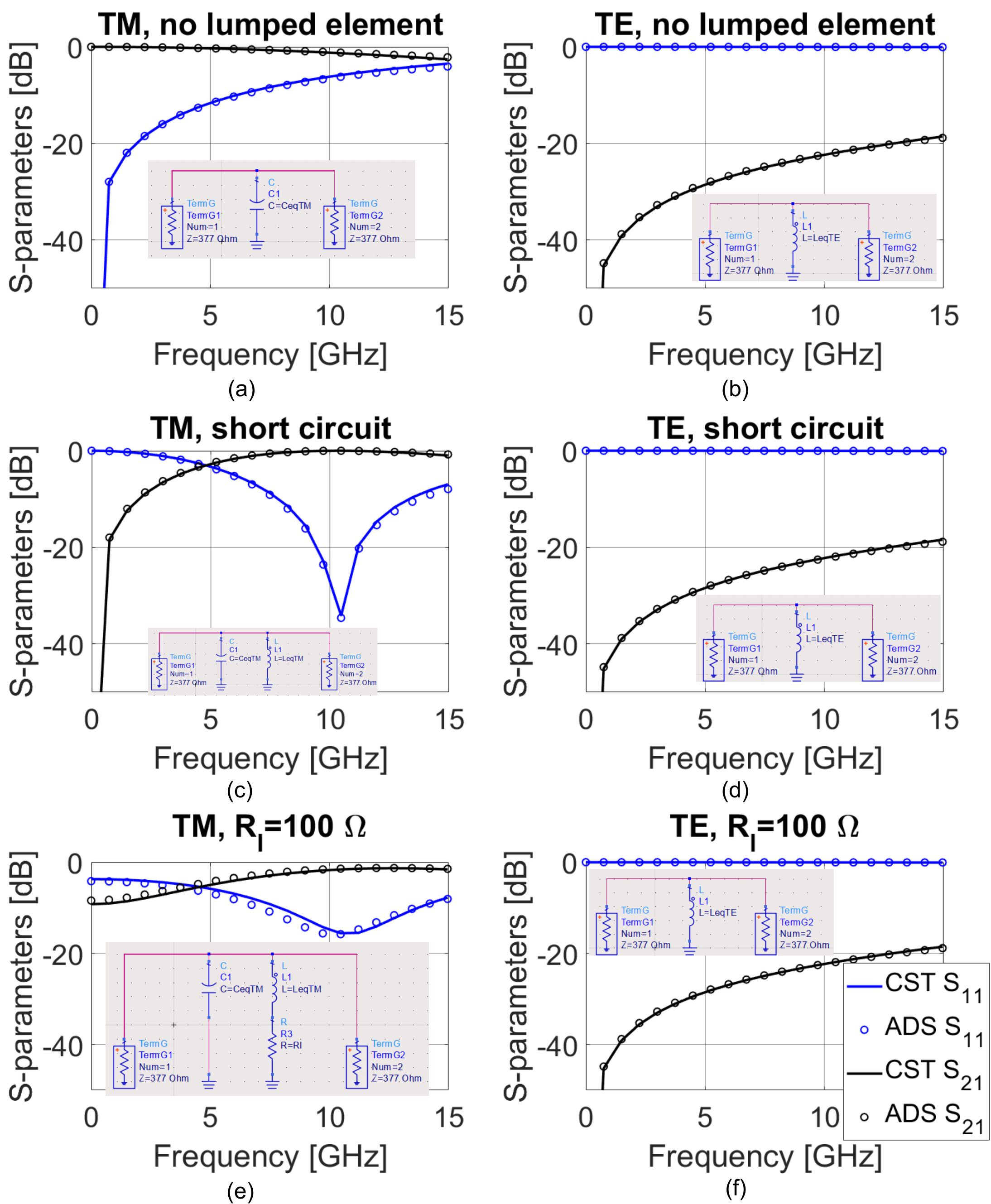}}
        \caption{S-parameters of the CST simulation and the proposed equivalent model for (a)-(b) an open,   (c)-(d) a short-circuited,   and (e)-(f) a   
        resistor-loaded metasurface. Parameters: $p_x=p_y=10$ mm, $w_y=3$ mm, $C_{\text{eq}}^{\text{TM}}=47$ fF, $L_{\text{eq}}^{\text{TE}}=230$ pH, $L_{\text{eq}}^{\text{TM}}=5.08$ nH, $R_l=100$ $\Omega$.}
        \label{fig:CSTMario}
\end{figure}

In~\cite{Berral2012}, a broadband analytical circuit model was proposed for the 1D grating without lumped elements. Unlike most previous models, the circuit approach captured the phenomenology associated with grating-lobe excitation. Simply, for TM incidence, the grating-discontinuity effect was represented by a shunt capacitor $C_{\text{eq}}^{\text{TM}}$. This element turned into an inductor $L_{\text{eq}}^{\text{TE}}$ for TE incidence. To validate this first approximation, which is the baseline for the proposed approach, a simulation of the structure in Fig.~\ref{Fig1} is carried out in the full-wave simulator CST Studio Suite, with $p_x=p_y=10$ mm, $w_y=3$ mm, considering 2 Floquet modes in each port (one for TM incidence and one for TE). The results of the simulation, in terms of S-parameters, are depicted in Fig.~\ref{fig:CSTMario}(a)~and~(b). In addition, the results of the equivalent circuit obtained from a Keysight Advanced Design System (ADS) fitting are shown. The values obtained for the capacitor and inductor are $C_{\text{eq}}^{\text{TM}} = 47\,$fF and $L_{\text{eq}}^{\text{TE}} = 230\,$pH, respectively.

Once the results for the first approximation have been obtained, the next step is to perform a simulation in which a tiny wire is added, joining the two extremes of the slit in the unit cell. This case could be interpreted as a lumped element (an ideal 0~$\Omega$ resistor or short-circuit, without inductive behavior) that has been added to the structure. Results for TM and TE incidences are depicted in Fig.~\ref{fig:CSTMario}(c)~and~\ref{fig:CSTMario}(d). As seen, a resonance has appeared for the TM incidence case. This resonance admits being modeled in the equivalent circuit as a shunt inductor, $L_{\text{eq}}^{\text{TM}}$, whose value can be obtained through the simulation as
\begin{equation}
    L_{\text{\text{eq}}}^{\text{TM}}=\frac{1}{C_{\text{eq}}^{\text{TM}}(2\pi f_r)^2},
\end{equation}
being $f_r$ the frequency where the resonance appears. In this case, the achieved value is $L_{\text{eq}}^{\text{TM}} = 5.08\,$nH. The result obtained from the equivalent circuit corresponds to the electromagnetic simulation performed. This is an important detail since it is shown that adding a lumped element between the slits causes them to no longer have the pure capacitive behavior expected for an isolated 1D grating illuminated by a TM wave \cite{Berral2012}. An inductive contribution appears as a consequence of the system's evolution from a 1D to a 2D periodic distribution. On the other hand, it can be seen that adding the infinitesimal wire when there is a TE incidence does not imply a change in the behavior of the grating. This is because the electric field is the same at both ends of the wire, so the electric currents will have the same value and direction at both ends, making the lumped element irrelevant to the circuit's behavior. 

To validate the conclusions obtained, the last simulated case consists of placing a 100 $\Omega$ resistor, $R_{\text{l}}$, in the slit. In this case, the resistor will replace the short circuit previously imposed for the TM incidence case, connecting it in series with the inductor, $L_{\text{eq}}^{\text{TM}}$. As seen in Fig.~\ref{fig:CSTMario}(e), the equivalent circuit fits perfectly the CST simulation, in which the resistor is placed as a lumped element. As anticipated, the TE incidence (Fig.~\ref{fig:CSTMario}(f)) is not affected by the resistor. This case is, therefore, discarded for the rest of the work, focusing only on TM incidence.

%%%%%%%%%%%%%%%%%%%%%%%%%%%%%%%%%%%%%%%%%%%%%%%%%%%%%%
\subsection{Isolated Metasurface}

As detailed in Sec. II.A, the circuit topology shown in \Fig{Fig1} captures perfectly the phenomenology associated with wave scattering in a metasurface loaded with a generic lumped element. Interestingly, both the capacitive term ($C_{\text{eq}}^{\text{TM}}$) and the inductive term ($L_{\text{eq}}^{\text{TM}}$), can be computed in a purely analytical form, by isolating the metasurface from the lumped element.

%\cite{Berral2012, Berral2012_1}, 2D \cite{Berral2015, Mesa2018, Dubrovka2006, Mesa2014, Alex2021}

The rationale for the analysis of the isolated metasurface is based on previous works that describe wave scattering in 1D and 2D metagratings \cite{Berral2012,  Berral2015, Mesa2014, Alex2021}, and, more recently, 3D \cite{Carlos_MTT3D, Alex3D_2023} metagratings. According to \cite{Berral2015}, the TM and TE admittances associated to the \emph{isolated} 2D metasurface can be computed as
\begin{equation} \label{YTM}
    Y_{\text{TM}}^{\text{TM}} = \sum_{\forall n,m \neq 0,0} N_{nm}^\mathrm{TM} \left(Y_{nm}^\mathrm{TM, (1)} + Y_{nm}^\mathrm{TM, (2)} \right)\, ,
\end{equation}
\begin{equation} \label{YTE}
    Y_{\text{TE}}^{\text{TM}} = \sum_{\forall n,m \neq 0,0} N_{nm}^\mathrm{TE} \left(Y_{nm}^\mathrm{TE, (1)} + Y_{nm}^\mathrm{TE, (2)} \right)\, ,
\end{equation}
where the superindex TM in $Y_{\text{TE}}^{\text{TM}}$ and $Y_{\text{TM}}^{\text{TM}}$ denotes the type of incidence, and the subindexes TE/TM denotes the nature of the Floquet harmonics taking place on the expression. We can clearly indentify $Y_{\text{TM}}^{\text{TM}}$ with $C_{\text{eq}}^{\text{TM}}$ and $Y_{\text{TE}}^{\text{TM}}$ with $L_{\text{eq}}^{\text{TM}}$. 

The elements $N_{nm}^\mathrm{TM/TE}$ are terms that identify the complex transformers of the $nm$-th TM/TE Floquet mode, which take the value
\begin{align}
& N_{n m}^{\mathrm{TM}}=\frac{k_{y m}^2}{k_{x n}^2+k_{y m}^2}\left|\frac{\tilde{E}_a\left(k_{x n}, k_{y m}\right)}{\widetilde{E}_a(0,0)}\right|^2 \, , \\
& N_{n m}^{\mathrm{TE}}=\frac{k_{x n}^2}{k_{x n}^2+k_{y m}^2}\left|\frac{\tilde{E}_a\left(k_{x n}, k_{y m}\right)}{\widetilde{E}_a(0,0)}\right|^2,
\end{align}
with 
\begin{multline} \label{Ea}
    \widetilde{E}_a(k_{xn}, k_{ym}) = \left[J_0\left(\frac{w_x}{2}\left|k_{x n}+\pi / w_x\right|\right) \right.
    \\+
    \left. J_0\left(\frac{w_x}{2}\left|k_{x n}-\pi / w_x\right|\right)\right] 
    \times \frac{\sin \left(\frac{w_y}{2} k_{y m}\right)}{k_{y m}} \, ,
\end{multline}
describing the Fourier transform of a suitable spatial profile (basis function) for a rectangular aperture of dimensions $w_x \times w_y$ under oblique TM illumination \cite{Rengarajan2005}. 

The admittances of the $nm$-th harmonic are computed as
\begin{align}
Y_{nm}^{\text{TM}, (i)} &= \frac{k_{0}\sqrt{\varepsilon_r^{(i)}}}{\eta^{(i)} \beta_{nm}^{(i)}}\, , \\
Y_{nm}^{\text{TE},(i)} &= \frac{\beta_{nm}^{(i)}}{\eta^{(i)} k_{0} \sqrt{\varepsilon_r^{(i)}}}\, ,
\end{align}
with  $\eta^{(i)}$ being the impedance of the input ($i=1$) and output ($i=2$) media and $k_0 = \omega/c$. The propagation constant   $\beta_{nm}^{(i)}$ reads
\begin{equation}
    \beta_{nm}^{(i)} = \sqrt{\varepsilon_r^{(i)}k_0^2 - k_{nm}^2}\, ,
\end{equation}
 where $k_{nm} = \sqrt{k_{xn}^2 + k_{ym}^2}$ with $k_{xn} = k_{x0} + 2\pi n /p_x$, $k_{ym} = k_{y0} + 2\pi m /p_y$, $k_{x0} = k_0 \sin \theta \cos \phi$, and $k_{y0} = k_0 \sin \theta \sin \phi$.

%--------------------------------------------------
\subsection{Analytical Circuit Model including Lumped Elements}

According to the circuit model shown in \Fig{Fig1}, the reflection coefficient of the 2D grating loaded with a lumped element with admittance $Y^\mathrm{Lump}$ is
\begin{equation} \label{R}
    R = \frac{Y_{00}^\mathrm{(1)} - Y_{00}^\mathrm{(2)}  - Y_\mathrm{tot}}{Y_{00}^\mathrm{(1)}  + Y_{00}^\mathrm{(2)}  + Y_\mathrm{tot}} \,
\end{equation}
where $Y_{00}^\mathrm{(1)/(2)}$ are the input/output transmission lines corresponding to the incident harmonic. Since TM incidence is considered, 
\begin{equation}
Y_{00}^{(1/2)} = \frac{\sqrt{\varepsilon_{\text{r}}^{(i)}}}{\eta^{(i)}\cos(\theta)} \hspace{5mm} i = 1,2\,.
\end{equation}
The transmission coefficient is calculated as
\begin{equation}
    T = 1 + R \, .
\end{equation}
In \eqref{R}, $Y_\mathrm{tot}$ represents the equivalent admittance of the 2D grating loaded with lumped elements, whose expression is extracted from the circuit model  (\Fig{Fig1}) and takes the form 
\begin{equation} \label{Ytot}
    Y_\mathrm{tot} = Y_{\text{TM}}^{\text{TM}} + \frac{1}{\frac{1}{Y_{\text{TE}}^{\text{TM}}} + \frac{1}{Y^\mathrm{Lump}}} \, .
\end{equation}

%%%%%%%%%%%%%%%%%%%%%%%%%%%%%%%%%%%%%%%%%%%%%%%%%%%%%%%%%%%%%%%%%%%%%%%%%%%%
\section{Numerical Results \& Validation}

This section shows some numerical examples to validate the analytical approach. Results are tested with full-wave simulations in commercial software CST Studio Suite.

\subsection{Lumped RLC Tank. Oblique Incidence}

\begin{figure}[!b]             
        \centering	           \subfigure[]{\includegraphics[width= 0.65\columnwidth]{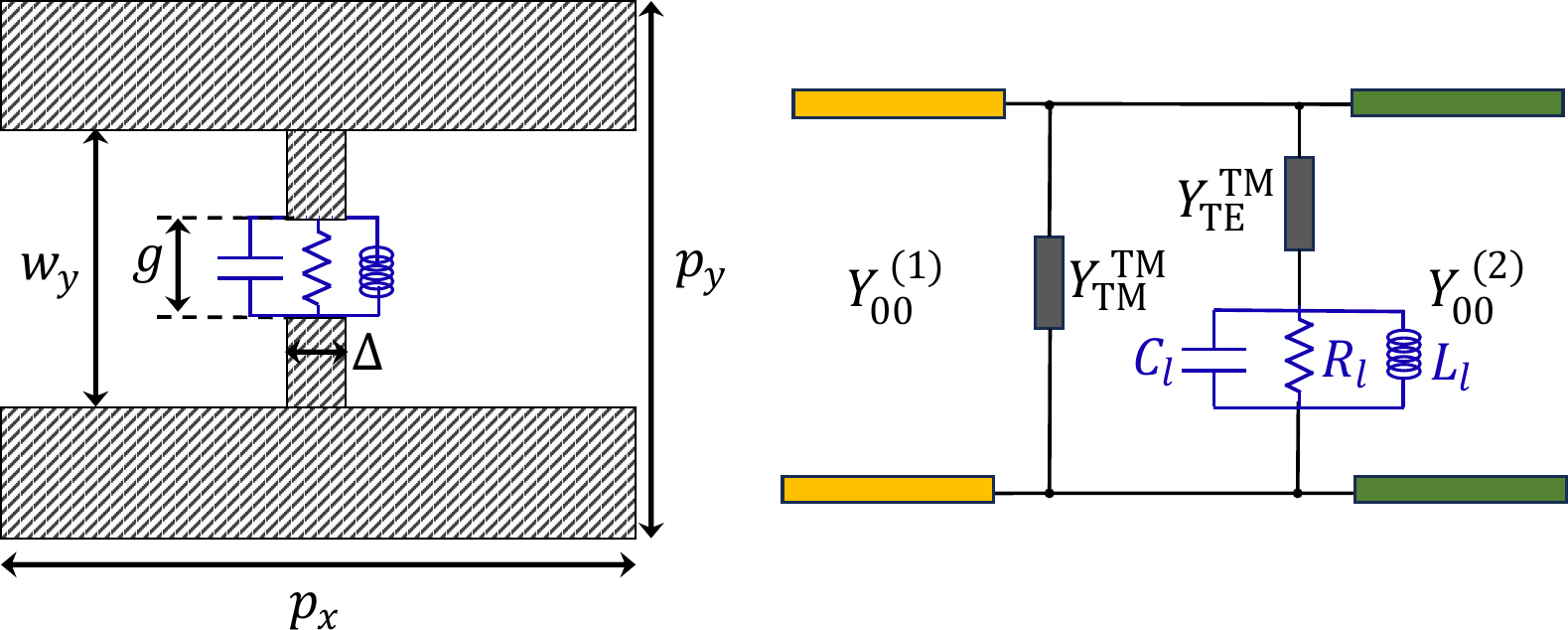}}
        \subfigure[]{\includegraphics[width= 0.51\columnwidth]{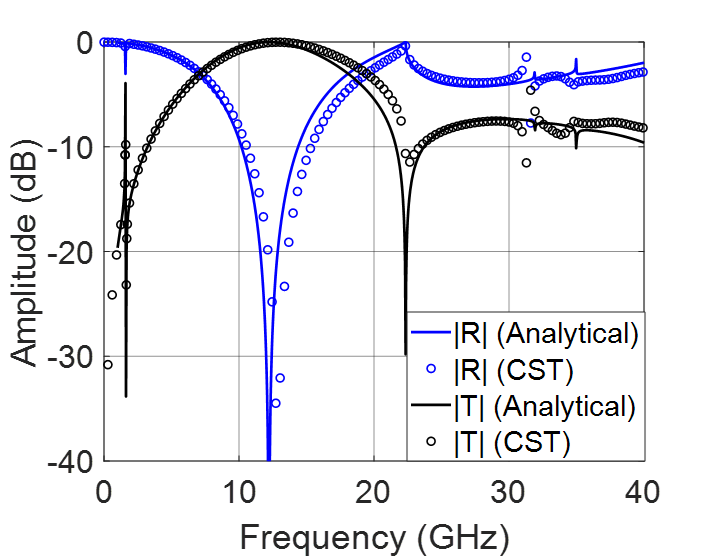}}
        \hspace*{-0.4cm}
        \subfigure[]{\includegraphics[width= 0.51\columnwidth]{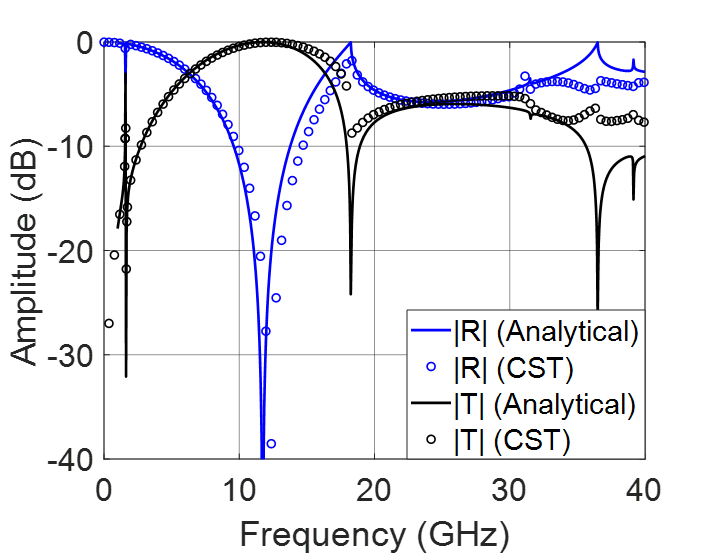}}
        \caption{(a) Schematic of 2D grating loaded with a lumped RLC tank. Amplitude of the reflection and transmission coefficient under TM oblique incidence: (b) $\theta=20^\mathrm{o}$,
         (c) $\theta=40^\mathrm{o}$.
        Parameters: $R_\mathrm{l} = 1000\, \Omega$, $C_\mathrm{l} = 100$ pF, $L_\mathrm{l} = 100$ pH, $\Delta = 0.1$ mm, $w_y = 3$ mm, $p_x = p_y = 10$ mm. }
        \label{RLCtank}
\end{figure}

The first numerical example, schematized in Fig. \ref{RLCtank}(a), includes a 2D metasurface loaded with a lumped RLC tank of values $R_l$, $C_l$ and $L_l$. This example is of general and complex nature, as oblique incidence ($\theta \neq 0$) is also tested. In the case of considering an RLC tank, the admittance of the lumped element to insert in eq. \eqref{Ytot} is $Y^\mathrm{Lump} = 1/R_\mathrm{l} + 1/(j\omega L_\mathrm{l}) + j\omega C_\mathrm{l}$. 

Figs. \ref{RLCtank}(b) and \ref{RLCtank}(c) present the amplitude of the reflection (blue curves) and transmission (black curves) coefficients for the oblique-incident angles $\theta=20^\mathrm{o}$ and $\theta=40^\mathrm{o}$, respectively. There is a good agreement between the analytical and CST results. As observed, the analytical circuit perfectly captures the frequency displacement of the grating lobe as the incident angle increases. The onset of the grating-lobe diffraction regime is given by the expression $f_{10} = c / (p  [1 +\sin \theta])$, which moves to 22.35 GHz for $\theta=20^\mathrm{o}$ and to 18.26 GHz for $\theta=40^\mathrm{o}$. Far beyond the onset of the diffraction regime, the analytical approach still provides accurate results, even for narrowband resonances above 30 GHz. The resonance below 2 GHz is created by the lumped RLC tank, which is well captured by the analytical circuit. 

The computation time of the analytical model is much lower than those of CST, as previously discussed. Especially the situation is aggravated when considering oblique incidence in the commercial simulator since more Floquet modes have to be included in the computation to obtain accurate results. Therefore, oblique incident scenarios notably increase computation times in CST while remaining rather similar whether normal or oblique incidence is considered in our analytical approach.

%---------------------------------------------------------------------
\subsection{Lumped Series RLC Resonator. Effect of $\Delta$ and $g$}

\begin{figure}[!t]             
        \centering	           
        \subfigure[]{\includegraphics[width= 0.5\columnwidth]{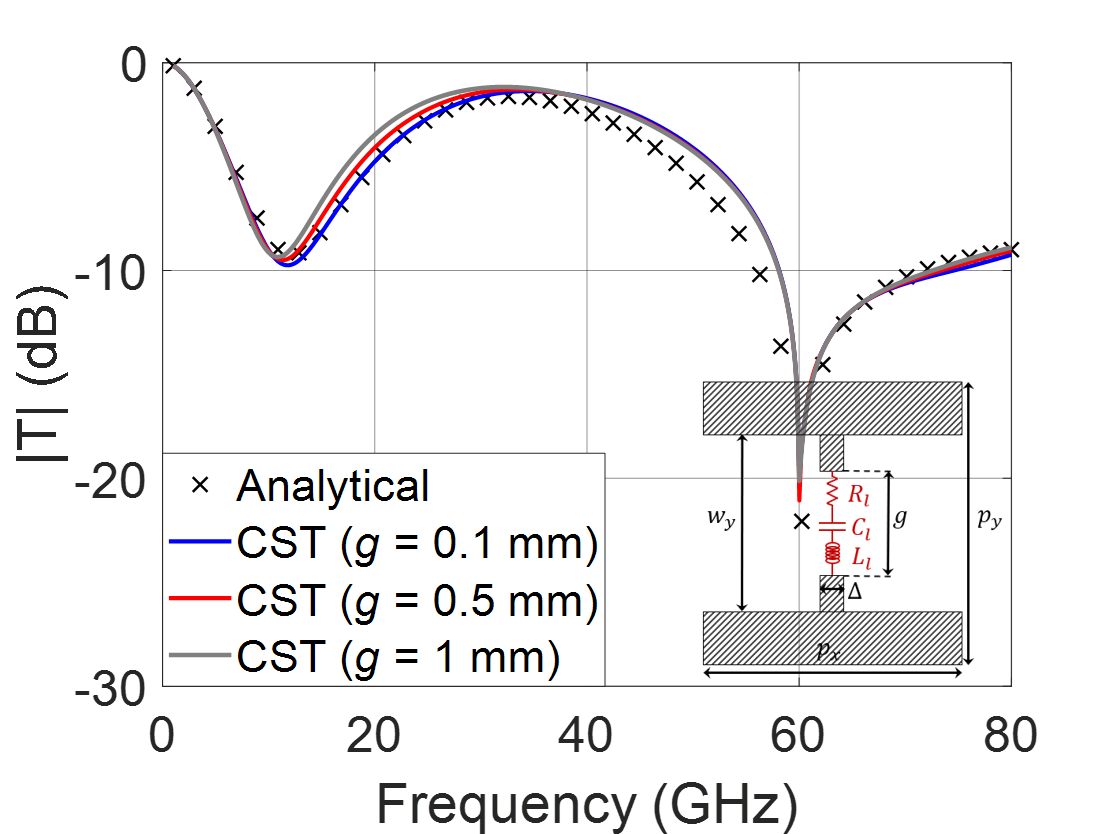}}
        \hspace*{-0.5cm}
        \subfigure[]{\includegraphics[width= 0.5\columnwidth]{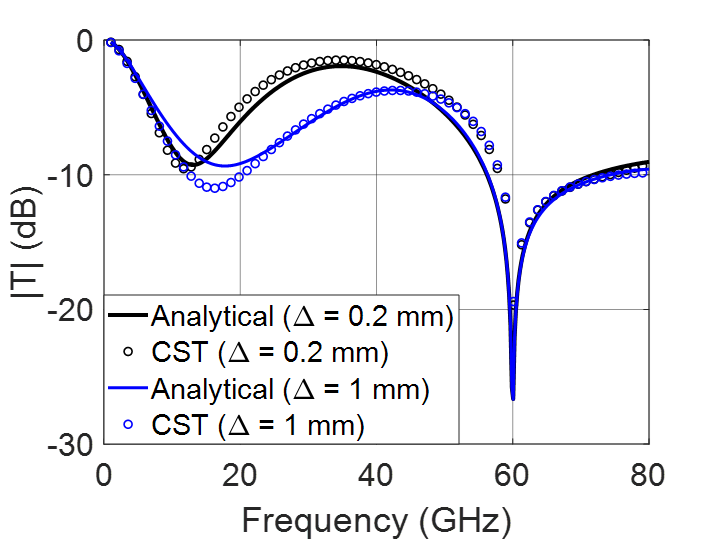}}
        \caption{Amplitude of the transmission coefficient $|T|$ in a 2D grating loaded with a lumped series RLC resonator series when varying: (a) the gap size $g$ ($\Delta = 0.1$ mm), and (b) the thickness of the strip $\Delta$ ($g=0.1$ mm).
        Parameters: $R_\mathrm{l} = 100\, \Omega$, $C_\mathrm{l} = 0.1$ pF, $L_\mathrm{l} = 0.1$ nH, $w_y = 2$ mm, $p_x = p_y = 5$ mm. TM normal incidence is assumed. }
        \label{RLCseries}
\end{figure}

Fig. \ref{RLCseries} illustrates the following numerical example as an inset. A different 2D metasurface is loaded with a lumped series RLC resonator of values $R_\mathrm{l}$, $C_\mathrm{l}$, and $L_\mathrm{l}$.
In the case of considering a series RLC resonator, the admittance of the lumped element is $Y^\mathrm{Lump} = \left[ (R_\mathrm{l} + j\omega L_\mathrm{l} + 1/(j\omega C_\mathrm{l}) \right]^{-1}$.

In this case, we perform a parametric study to see the effect of the thickness of the strip $\Delta$ and its gap size $g$ on the method's accuracy. Results are illustrated in Fig. \ref{RLCseries}. Only the transmission coefficient is plotted in this case to ease the visualization. Technically, the formulation developed in Sec. II assumes that the length of the gap is $g=0$. However, as Fig. \ref{RLCseries}(a) illustrates, increasing the gap size does not significantly affect the method's accuracy. Gap sizes of 0.5 mm and 1 mm are realistic values where a microwave lumped element (varactor, PIN diode, etc.) may be soldered. 

Then, Fig. \ref{RLCseries}(b) shows the effect of varying the strip thickness $\Delta$. The circuit's topology presented in \Fig{Fig1} ideally assumes that $\Delta \ll p_x$. Fig. \ref{RLCseries}(b) shows how the method's accuracy slightly deteriorates as the thickness $\Delta$ increases. However, good agreement is still observed with the commercial software. Note that, from a practical point of view, thicknesses $\Delta \geq 0.1$ mm are easily manufacturable in microstrip or similar technologies. The combined effect of increasing the gap $g$ and the strip thickness $\Delta$ might deteriorate the analytical approach results.

%-----------------------------------------------------------------
\subsection{Lumped Varactor Diode}

\begin{figure}[!t]             
        \centering	              \subfigure{\includegraphics[width= 0.9\columnwidth]{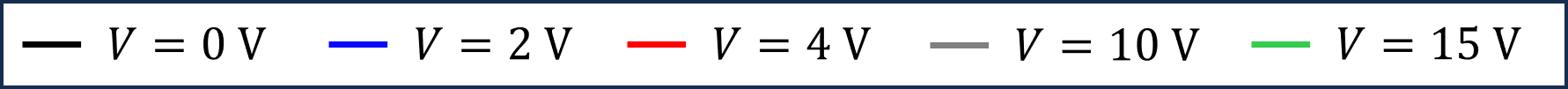}} 
        \vspace{-0.4cm}\\
    \setcounter{subfigure}{0}
        \subfigure[]{\includegraphics[width= 0.48\columnwidth]{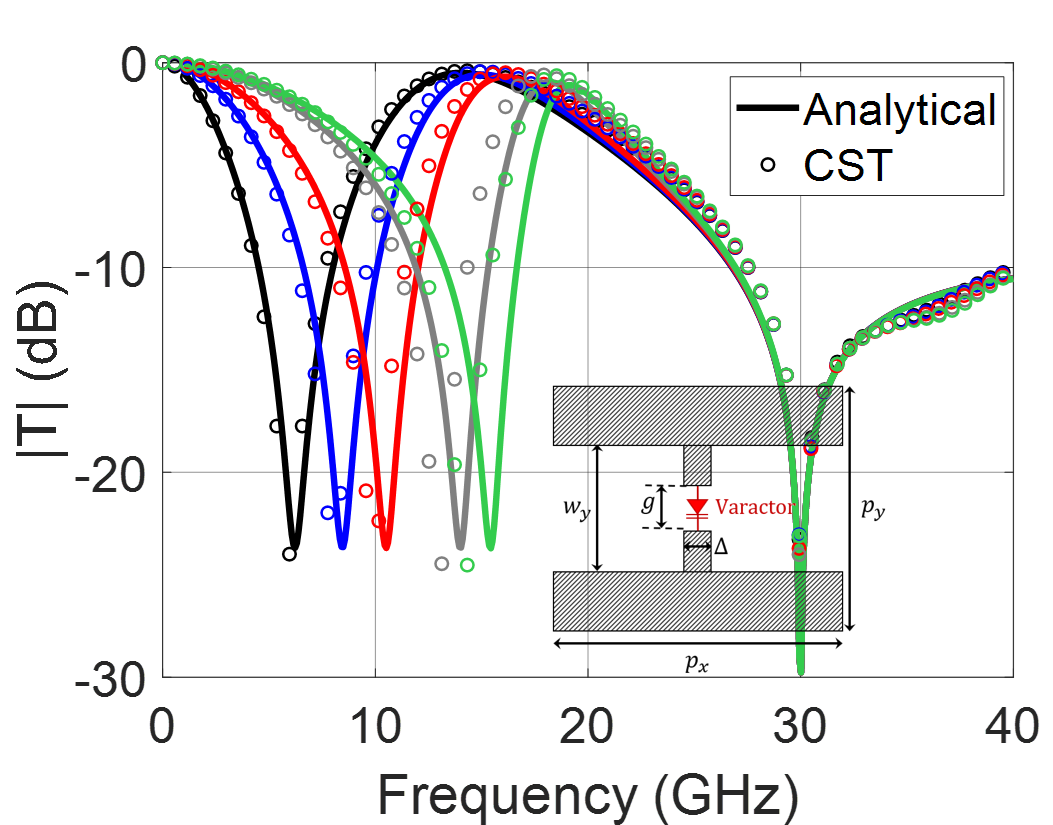}}\
        \hspace*{-0.3cm}
        \subfigure[]{\includegraphics[width= 0.48\columnwidth]{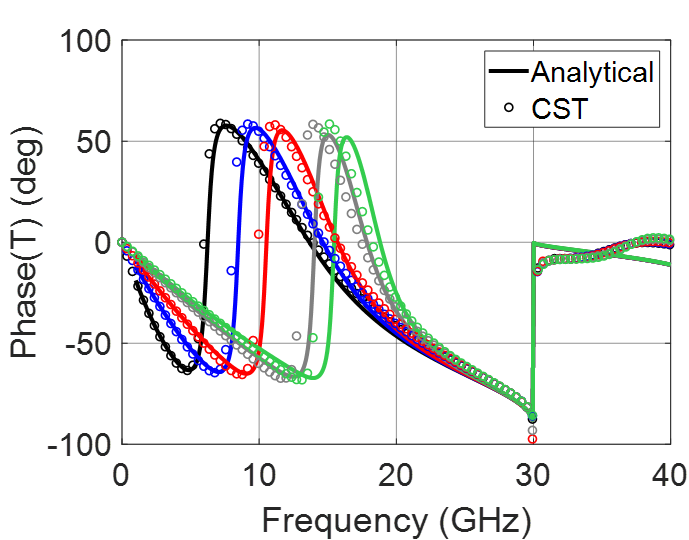}}\
        \caption{Transmission in a metasurface loaded with a varactor: (a) amplitude, (b) phase. 
        Parameters:  $w_y = 3$ mm, $p_x = p_y = 10$ mm. TM incidence. }
        \label{varactor}
\end{figure}

Now, we consider the case where the metasurface is loaded with a varactor diode [see Fig. \ref{varactor}(a)]. This is a common scenario, highly interesting and useful in the design of reconfigurable devices, such as reflectarrays/transmitarrays and time-modulated modules, which have been gaining attention lately \cite{JSGomez2019, metalumped2, Tretyakovlumped, Wu2020}. We consider the commercial microwave varactor MAVR-011020-1411 \cite{varactordiode} for this example. At low frequencies, the varactor diode can be modeled as a series RC circuit, with $R_\mathrm{l} = 13.2\, \Omega$ and a varying capacitor $C_l$ controlled by a bias voltage $V$. The relation between the applied $V$ and the obtained  $C_\mathrm{l}$ is as follows: $V = \{0, 2, 4, 10, 15 \}$ V $\leftrightarrow C_\mathrm{l}= \{0.233, 0.125, 0.080, 0.0439, 0.0357 \}$ pF. 

Figs. \ref{varactor}(b) and \ref{varactor}(c) illustrate the amplitude and phase terms of the transmission coefficient, respectively. As observed, there is a good correspondence between the analytical results and those obtained by CST in a wideband range of frequencies, even above the diffraction regime (30 GHz). An increase in the reverse bias voltage $V$, is translated into a decrease in the series capacitance $C_\mathrm{l}$ and, subsequently, a decrease in the lumped admittance $Y^\mathrm{Lump} = \left[ (R_\mathrm{l} + 1/(j\omega C_\mathrm{l}) \right]^{-1}$. Therefore, lower current flows through the varactor branch so that the TM (capacitive) term clearly predominates over the TE (inductive) one, increasing the transmission level at the low frequencies (near DC). In other words, the zero-transmission peak shifts to higher frequencies as the bias voltage increases and $C_\mathrm{l}$ is smaller. This example highlights the physical insight that the analytical model provides about the scattering phenomenon.

%-----------------------------------------------------------------
\section{Realistic Design}

\begin{figure}[!t]             
        \centering	           
        \subfigure[]{\includegraphics[width= 0.6 \columnwidth, height = 0.32\columnwidth]{Absorbente.pdf}}
        \hspace*{-0.5cm}
        \subfigure[]{\includegraphics[width= 0.75\columnwidth]{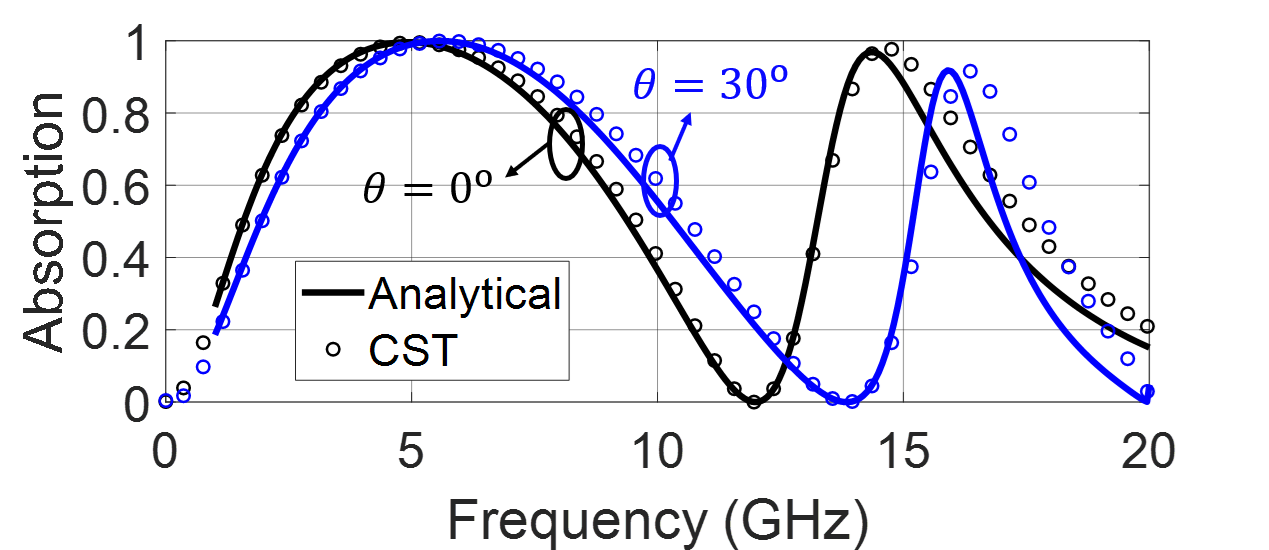}}
        \caption{(a) Sketch of a resistor-based meta-absorber. (b) Absorption under TM normal (black) and oblique (blue) incidence. Solid lines and circles represent the analytical and CST results. Parameters: $R_\mathrm{l} = 310\, \Omega$, $\Delta = 0.1$ mm, $g = 0.5$ mm, $w_y = 3$ mm, $p_x = p_y = 10$ mm, $d = 12.5$ mm.  }
        \label{Absorbente}
\end{figure} 

The analytical model can also be used as an efficient and physically-insightful \emph{design} tool. As an example, we apply the analytical framework for the
design of a dual-band absorber. Fig. \ref{Absorbente}(a) illustrates the proposed device, based on a metasurface loaded with a lumped resistor $R_\mathrm{l}$. To achieve a higher absorption ratio, the resistor-loaded metasurface is backed by a metallic plate, separated by a distance, $d$. The medium separating the two is air, although other dielectric materials could be chosen. The analytical formulation stays identical, except for the term $Y_{nm}^{\text{TM/TE}, (2)}$ used in Eqs. \eqref{YTM}, \eqref{YTE} and \eqref{R}, that should be replaced by $Y_{\mathrm{in}, nm}^{\text{TM/TE}, (2)} = -jY_{nm}^{\text{TM/TE}, (2)} \cot(\beta_{nm}^{(2)}d)$.  

%Nevertheless, this metasurface could be manufactured with foam-like materials.

Fig. \ref{Absorbente}(b) presents the absorption parameter. A greater absorption rate is obtained near the $\lambda/4$ and $3\lambda/4$ resonances (around $6$ GHz and $18$ GHz). Under normal TM incidence, a fractional bandwidth (FBW) of $72.8$$\%$ is achieved in the first band (from $3.28$ GHz to $7.04$ GHz), assuming an absorption of more than $90$$\%$. A second absorption band appears from $14.08$ GHz to $15.38$ GHz, giving a FBW of $8.8$$\%$. This second band can still be used, as the first grating lobe is excited at $30$ GHz. Under oblique TM incidence ($\theta=30^{\text{o}}$), the first grating lobe appears around $20$ GHz. Even so, the first absorption bandwidth obtains a FBW = $68$$\%$ (from $3.84$ GHz to $7.8$ GHz), showing robustness under oblique-incidence conditions.  In this case, CST took  1 minute and more than 8 minutes when normal and oblique incidence was considered, respectively. The analytical approach took less than 1 second in both cases.

%A greater absorption rate should be obtained near the $\lambda/4$ and $3\lambda/4$ resonances when the grounded transmission line behaves as an open circuit (around $6$ GHz and $18$ GHz). However, the resistor value has been chosen to optimize the fractional bandwidth obtaining a better matching of $Y_{\text{tot}}$ around $5.2$ GHz and $14.7$ GHz.

%%%%%%%%%%%%%%%%%%%%%%%%%%%%%%%%%%%%%%%%%%%%%%%%%%%%%%
\section{Conclusion}
In this paper, we have proposed a novel mathematical framework for analyzing metasurfaces loaded with lumped elements. The present approach is fully-analytical; thus, physical insight is given into complex scattering phenomena. Moreover, it notably reduces the computation time compared to state-of-art and commercial full-wave approaches. In normal and oblique incidence, the analytical framework works in a wide range of frequencies, even far beyond the onset of the grating-lobe regime. Finally, the analytical approach has been tested against full-wave numerical results in CST. Results include the insertion of a lumped (i) RLC tank, (ii) series RLC resonator, (iii) varactor diode, (iv) and a microwave absorber. Accurate results are obtained even for oblique incidences of $\theta=40^\mathrm{o}$.

%Moreover, we have performed a parametric study to test the effect of varying some dimensions in which the lumped element is soldered. 

\section*{Acknowledgments}
We would like to thank M. Fernández-Pantoja for using his Keysight Advanced Design System (ADS) license.

%BIBLIOGRAPHY
\bibliographystyle{IEEEtran}
\bibliography{IEEEabrv,ref}

\end{document}